\documentclass[10pt]{article}
\usepackage{ametsoc2col}
\newcommand{\myabstract}{
The MaddenÐJulian oscillation (MJO) is the dominant mode of intraseasonal variability in the tropics. Despite its primary importance, a generally accepted theory that accounts for fundamental features of the MJO, including its propagation speed, planetary horizontal scale, multi-scale features, and quadrupole structures, remains elusive. In this study, we use a shallow water model to simulate the MJO. In our model, convection is parameterized as a short-duration localized mass source, and is triggered when the layer thickness falls below a critical value. Radiation is parameterized as a steady uniform mass sink. Slowly eastward propagating (MJO-like) signals and red noise spectra are observed in our simulations. In the time-longitude domain, MJO-like signals with multi-scale structures are observed. In the Fourier domain, spectral peaks associated with the MJO-like signals are observed. In the longitude-latitude map view, quadrupole vortex structures associated with the MJO-like signals are observed. We propose that the simulated MJO signal is an interference pattern of  westward and eastward inertia-gravity(WIG and EIG) waves. Its propagation speed is one half of the speed difference between the WIG and EIG waves. The horizontal scale of its large-scale envelope is determined by the bandwidth of the excited waves, and the bandwidth is controlled by number density of convection events. Our results suggest that the MJO perhaps is not a large-scale low-frequency wave, in which convection acts as quasi-equilibrium adjustment. Small-scale high-frequency waves might be important.}
\begin{document}
%
%
\title{\textbf{\large{Triggered convection, gravity waves, and the MJO: A shallow water model}}}
%
%
\author{\textsc{Da Yang}
				\thanks{\textit{Corresponding author address:} 
				Da Yang, Division of Geological and Planetary Sciences, MC 150-21, California Institute of Technology, Pasadena, CA 91125, USA. 
				\newline{E-mail: dyang@caltech.edu}}\quad\textsc{and Andrew P. Ingersoll}\\
\textit{\footnotesize{Division of Geological and Planetary Sciences, California Institute of Technology, Pasadena, CA, USA.}}
}
%
\ifthenelse{\boolean{dc}}
{
\twocolumn[
\begin{@twocolumnfalse}
\amstitle

\begin{center}
\begin{minipage}{13.0cm}
\begin{abstract}
	\myabstract
	\newline
	\begin{center}
		\rule{38mm}{0.2mm}
	\end{center}
\end{abstract}
\end{minipage}
\end{center}
\end{@twocolumnfalse}
]
}
{
\amstitle
\begin{abstract}
\myabstract
\end{abstract}
\newpage
}
\section{Introduction}

The Madden-Julian Oscillation (MJO) is the dominant intraseasonal variability in the tropical atmosphere. It is a planetary scale, slowly eastward propagating (about 5 $m\ s^{-1}$) perturbation of both dynamical and thermodynamical fields. During an MJO event, a positive convection and rainfall anomaly develops in the western Indian Ocean and propagates to the western Pacific Ocean. Once the perturbation reaches the date line, the perturbation is largely confined to dynamical fields. The associated planetary-scale wind structure is baroclinic, and is characterized by low level convergence and upper level divergence (\cite{MaddenJulian1972}; \cite{MaddenJulian1994}; \cite{Hendon1994}). More details can be found in comprehensive reviews, such as, \cite{mj05} and \cite{zhang2005}. In addition, within the MJO envelope, there are both westward and eastward moving fine scale structures (\cite{nakazawa1988}, \cite{HendonLiebmann1994}). \\

The significance of the MJO for phenomena such as monsoon onset, ENSO, and rainfall patterns in the Tropics has been well established (\cite{zhang2005}). However, a successful MJO theory is elusive. For a historical and detailed review of theories, please refer to \cite{zhang2005} and \cite{wang05}. Here we will review three schools of theories. One school considers the MJO as a largeÐscale unstable mode in the tropics, and it is often referred to as the moisture mode. The moisture mode arises from positive feedbacks between precipitation and the source of moist static energy, e.g., \cite{NeelinYu1994}, \cite{Sobel2001}, \cite{FuchsRaymond2002, FuchsRaymond2005}, \cite{Bretherton2005}, \cite{FuchsRaymond2007}, \cite{Maloney2009}, \cite{Raymond09}, \cite{AndersenKuang2012}. Another school still considers the MJO is a large-scale mode, but the major instability to maintain the planetary scale envelope happens in the synoptic scales. Both observational and modeling studies suggest that high-frequency small-scale waves are important to the MJO, e.g., \cite{HendonLiebmann1994}, \cite{TungYanai2002I}, \cite{Moncrieff2004}, \cite{Biello2005}, \cite{Khouider2012CMT}. Thus \cite{MajdaStechmann2009, MajdaStechmann2011} emphasize the importance of small-scale waves within the MJO envelope, but they parameterize the effect of the waves. In their model, it is only the amplitude of the wave activity envelope that is needed, not any of the details of the particular synoptic scale waves that make up the envelope. The third school considers the MJO as a wave packet of a certain type of equatorial waves, and the MJO propagates with the group velocity of the equatorial waves. \cite{YangIngersoll2011} hypothesize that the MJO is a mixed Rossby-gravity (MRG) wave packet that propagates with the MRG group velocity, and they test this hypothesis with both an idealized GCM and with the Outgoing Longwave Radiation dataset. They force with a westward-moving heat source that lasts for 10 or 20 days. At the same time, \cite{Solodoch2011} suggest that the MJO could be a forced response to the MRG wave group in the quasi-equilibrium (QE) convection and wind-induced surface heat exchange context. Neither theory considers inertia-gravity (IG) waves, which have frequencies greater than 0.5 cycles day$^{-1}$. In the QE context, these high-frequency small-scale waves will be damped the fastest, and it is possible that the QE approach does not apply there. Therefore the role of high-frequency small-scale waves has not really been investigated in any of the theories, including the wave packet theories of the MJO. \textbf{Here we present a simple MJO model with explicit triggered (non QE) convection that emphasizes the multi-scale structures of the MJO}.\\

General circulation models (GCMs) simulate tropical intraseasonal variability with varying degrees of fidelity. Intercomparison studies (\cite{Lin2006}) show that most GCMs underestimate the MJO variance. The weak MJO signals in GCMs are believed due to inadequate convection schemes in the GCMs. \cite{Holloway12} compare limited-area simulations of the tropical atmosphere over a very large domain at different horizontal resolutions with both parameterized and explicit convection versions for a 10 day MJO case study in April 2009. They claim that the parameterized models consume CAPE and reach radiative-convective equilibrium too quickly, and lack the ability to transition from suppressed to active conditions and vice versa. As a result, propagating MJO signals are absent in the parameterized runs. \\

It is widely believed that improvements in the representation of subgrid-scale processes in the model would lead to a more accurate MJO depiction. However, many of conventional convection parameterizations are based on the quasi-equilibrium (QE) idea, which tends to keep the temperature profile close to the moist adiabat, and tends to damp small-scale waves. A major weakness of parameterizations is that they artificially separate subgrid-scale processes (departures from QE) from the large scale processes, although the small- and large-scale processes interact in nature. \\

Recent studies show improved MJO simulations with superparameterized (SP) Community Atmosphere Models (CAM) (\cite{BenedictRandallSPCAM09}). The SP-CAM replaces conventional boundary layer, moist convection parameterizations with a cloud-resolving model embedded in each CAM grid cell (\cite{KhairoutdinovRandall2001}). The SP-CAM deals with sub-grid scale variability more accurately. Our interpretation of the SP-CAM results is that the cloud-resolving model excites high-frequency small-scale waves, which are crucial to the MJO.\\

There are two common ways of treating convection: triggered and statistical equilibrium convection (or QE). There is not a clear distinction between these two categories. Over a long period, in the tropics, the generation of convective available potential energy (CAPE) by large-scale processes nearly balances its consumption by convection. The convection may be considered to be in a state of statistical equilibrium with the large-scale circulation. It is first applied by \cite{arakawa1974}.  This idea has been validated using observational data sets by \cite{XuEmanuel89} and \cite{Holloway07}. QE is a very good assumption to study large-scale circulations that vary slowly with time compared with convective timescales. Such circulations include tropical cyclones, the Hadley cell and monsoon circulations ($e.g.$, \cite{Emanuel1994}, \cite{Emanuel2007}),  but so far the MJO has not been successfully simulated under the QE context. If we are interested in high-frequency small-scale waves, however, QE does not work well. First, over a short period, CAPE builds up. When some threshold is reached, convection is triggered and CAPE is released. Second, in a QE scheme, convection will damp small-scale waves faster and leave the large-scale waves. We are investigating triggered convection in order to simulate high-frequency small-scale waves.\\

Motivated by the success of SP-CAM simulations, we develop a theory emphasizing the role of high-frequency small-scale waves. In this paper, we use a shallow water model with triggered convection and radiation represented as Newtonian relaxation. Slowly eastward propagating (MJO-like) signals are observed in our simulations. Instead of thinking of the MJO as a large-scale stable or unstable mode, we propose the MJO-like signals are interference patterns of westward and eastward inertia gravity (WIG and EIG) waves. Eastward propagation is due to the zonal asymmetry of the EIG and WIG waves. In section 2, we will introduce the shallow water model used in this study. In sections 3 and 4 we show our simulation results and attempts to understand the simulated signal. In section 5, we will discuss our results, and present our conclusions and future work.\\

\section{Model description}
Shallow water model is a two-dimensional model. We simulate the upper troposphere by assuming the first baroclinic mode, since the large-scale circulation associated with the MJO shows the first baroclinic structure. Thus divergence in the model refers to upper level divergence and low level convergence. Similarly, large layer thickness in the model corresponds to high pressure aloft and low pressure near the surface. In this section, we will introduce the shallow water model used in this study.\\

This shallow water model describes the evolution of constant density, incompressible fluid over the surface of the sphere. The model equations are:

\begin{align}\label{2dU} 
   \partial_{t}u = fv - \frac{u}{acos\theta}\partial_{\lambda}u + \frac{v}{a}\partial_{\theta}u + \frac{uvtan\theta}{a} - \frac{1}{acos\theta}\partial_{\lambda}\phi,
\end{align}
\begin{align}\label{2dV}
   \partial_{t}v = -fu - \frac{u}{acos\theta}\partial_{\lambda}v + \frac{v}{a}\partial_{\theta}v + \frac{u^2tan\theta}{a} - \frac{1}{a}\partial_{\theta}\phi, 
\end{align}
\begin{align}\label{2dh}
   \partial_{t}\phi = -\nabla\cdot(\vec{V}\phi) + s, 
\end{align}
\begin{align}\label{2dS}
   s = q - r.
\end{align}

Equations \eqref{2dU} and \eqref{2dV} are momentum equations, where u and v are zonal and meridional velocities; $\phi$ represents geopotential corresponding to the layer thickness; $a$ is the earth radius; $f$ is the Coriolis parameter (a.k.a. planetary vorticity); $\lambda$ and $\theta$ represent longitude and latitude in radians. Equation \eqref{2dh} is the continuity equation, and $s$ includes both mass sources and sinks. In Eq. \eqref{2dS}, $q$ represents convective heating, which is a mass source; and $r$ represents radiative cooling, which is a mass sink.\\

In this model, convection events are triggered by a low value of the layer thickness, i.e. if the layer thickness is lower than a threshold $\phi_{c}$, convection will start to add mass into this shallow water system:

\begin{align}\label{conv}
   q =  
        \begin{cases}
        \frac{q_{0}}{\tau_c A_o}(1 - (\frac{\triangle t - \tau_c/2}{\tau_c/2})^2)(1-\frac{L^2}{R^2}) \quad \quad when\ \phi < \phi_c, \ 0 < \triangle t < \tau_c,\ and\ L^2 \leqslant R^2,\\
        0      \qquad  \qquad  \qquad  \qquad  \qquad           \qquad               \qquad                                                            otherwise,
        \end{cases}
\end{align}
where $q_{0}$ is a free parameter of the heating amplitude, and $\tau_c$ is the convective timescale. Each convective event operates in a certain area $A_o = \pi R^2$, where $R$ is the radius of each convective event. $L = \sqrt{\triangle x^2 + \triangle y^2}$ measures the distance from the convective center. Radiative cooling is constant in both time and space. The ratio $\frac{3r}{q_o}$ determines N, the average number of convection events per unit area, per unit time. In a statistically steady state, the total mass of this system will not change with time. Convection, the mass source, will be balanced by the mass sink, radiation. The equilibrium geopotential is $\sim$ $\phi_c$.  \\

The SW equations have characteristic length and time scales through the planetary radius and the rotation. In addition to the planetary radius and rotation, there are four parameters: the equilibrium geopotianl $\phi_c$, the area of one convection event $A_o$, the convective timescale $\tau_c$, and the number density of convection events N, which is controlled by $r$ through the relation $N \sim \frac{3r}{q_o}$ . Both $q_o$ and $r$ are small, so that only the ratio between the two terms matters. \\

The Kelvin wave speed, $c$ is equal to $\sqrt \phi_c$. In this study, we fix $c \sim 16\ m\ s^{-1}$. The mean depth $h_e$ (a.k.a. equivalent depth)  of this SW system is given by $\frac{\phi_c}{g}$. Radius of convection $R$ is 3 degrees of latitude, which is approximately the size of the grid in T42 simulations. Parameter values of our control simulation are documented in table 1. We vary the horizontal resolution from T42 to T170, and find our main results are not sensitive to resolution. The results presented in this paper are mainly from T42 simulations. We solve these equations in spherical coordinates by using the spectral dynamical core of the Geophysical Fluid Dynamics LaboratoryÕs (GFDLÕs) Flexible Modeling System. Although we do not have a moisture variable explicitly in our model, our model does illustrate the importance of moisture to the MJO.\\

\section{Simulation results}

\begin{figure*}[htbp]
  \noindent\includegraphics[width=39pc,angle=0]{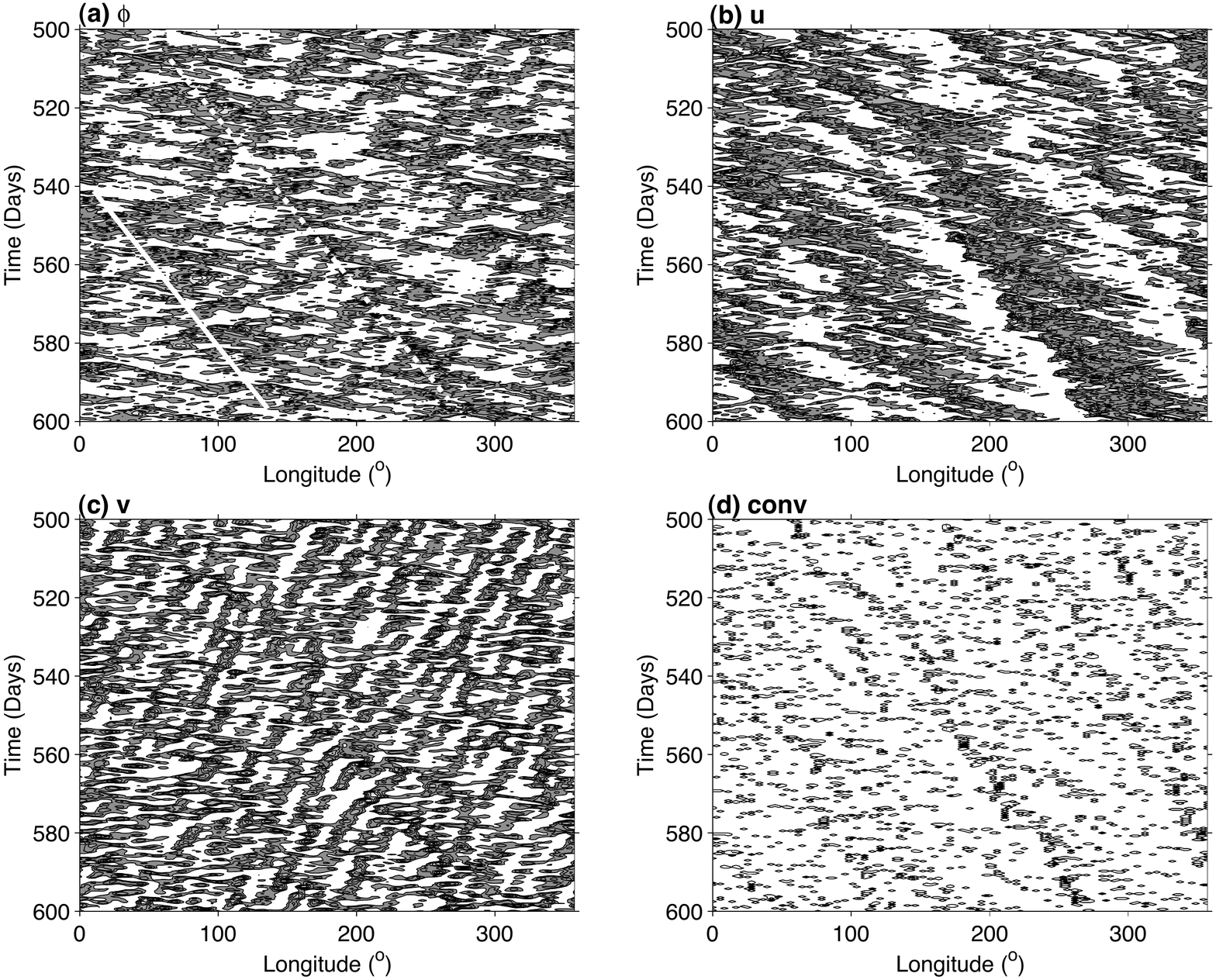}\\
  \caption{Hovmoller diagrams of the (a) geopotential, (b) zonal wind, (c) meridional wind and (d) convective heating of our shallow water simulation from -$15^{o}$ to $15^{o}$ latitude. The geopotential, zonal wind and convective heating are symmetric components about the equator, and the meridional wind is the anti-symmetric component. The white represents low, and the black represents high.}\label{hd2d}
\end{figure*}

Figure \ref{hd2d} shows Hovmoller diagrams of our shallow water simulation from day 500 to day 600. Figure 1 shows the symmetric components of the gepotential, zonal wind, and convective heating and the antisymmetric components of meridional wind. These symmetries and asymmetries are respect to the equator. Such meridional symmetry excludes the even meridional-wind modes and leaves the odd modes. Figure \ref{hd2d}a shows the geopotential. There are two major large-scale events labeled A and B. These are the MJO-like signals. They move eastward at $\sim$ 3.2 m s$^{-1}$. Small scale waves are present, and they include Kelvin waves, IG waves and Rossby waves. Since the IG waves are small scale waves, the absolute values of their speed are close to the Kelvin wave speed, which is $\sim$ 16 m s$^{-1}$.  Figure \ref{hd2d}b shows the zonal wind. The white represents westward zonal wind, and the black represents eastward zonal wind. The edge between the white and black indicates the divergence of zonal wind. Two regions of large scale divergence are observed, and they are collocated with events A and B in Fig. \ref{hd2d}a. Figure \ref{hd2d}c shows the meridional velocity v. Small-scale westward and eastward waves are observed, but large scale envelopes are not clear in the v field. Figure \ref{hd2d}d shows convective heating. Convection is a small-scale, short-period process, but two organized long-lasting events are observed, and they are collocated with events A and B in Fig. \ref{hd2d}a and with large-scale divergence in Fig. \ref{hd2d}b. Similar to the observation, large-scale divergence is collocated with convective centers, and dynamical fields are coupled to convection.\\

Figure \ref{hd2d}b is the Hovmoller diagram of the zonal wind from a T42 simulation. Figure 2 shows the Hovmoller diagrams of the zonal wind from T85 and T170 simulations with the same parameters. The size of the convection in kilometers is the same. They both show propagation speeds and horizontal structures that are  similar to the T42 simulations. This comparison suggests that the T42 simulations have already converged. In the rest of the paper, we will present the simulation results from the T42 simulations. \\
 
To understand the multi-scale structures in our simulation, we carried out space-time spectral analysis as pioneered by \cite{wk99} (hereafter WK99). Figures \ref{spectra}a and b show the power spectra of the symmetric and anti-symmetric components (about the equator) of the zonal wind. Superimposed curves represent dispersion relations of equatorial waves for $\sim 16\ m\ s^{-1}$  Kelvin wave speed. The dispersion curves of the equatorial waves are first derived by \cite{Matsuno1966}. Different equatorial waves are characterized by different dispersion relations and meridional mode numbers n. The spectral power under the superimposed curves is associated with the corresponding equatorial waves. In Fig. \ref{spectra}a, we can see spectral peaks associated with n = 1 Rossby, Kelvin and n = 1, 3 IG waves. Consistent with the slowly eastward moving signals in Fig. \ref{hd2d}, Fig. \ref{spectra}a has an MJO-like signal within the white box. This signal has planetary scale and low frequency. In the rest of this paper, we will try to understand this interesting phenomena. In Fig. \ref{spectra}b, we can see spectral peaks associated with n = 2 Rossby, MRG and n = 2 IG waves. A striking feature of Fig. \ref{spectra} is the intense power associated with high frequency IG waves. High IG wave activity is associated with the convective parameterization in our model. We will argue that high IG wave activity is the key to the MJO-like signal. Most of the power is concentrated in low wavenumber and low frequency region, except for the spectral peaks associated with high frequency IG waves. This is, in general, a red noise spectrum. Although different in details, Fig. \ref{spectra} captures some fundamental features of the observed spectra by WK99. \\
 
Figure \ref{map} shows the horizontal structure of the MJO-like signal in our simulation. To get Fig. \ref{map}, we first get the MJO filtered signal through zonal wind field in the wavenumber-frequency domain. The filtering excludes all wavenumbers and frequencies except those in the MJO box (Fig. \ref{spectra}a). Following \cite{WheelerHendon2004}, we carried out empirical orthogonal function (EOF) analysis of the MJO filtered signal. We found the first two EOFs can contribute $\sim$ 85\% of the total variance of the MJO filtered signal. The EOFs together with the corresponding principal components (PCs) show the propagation behavior of the MJO-like signal. Combining all the phases of the MJO-like signal, we get the horizontal structure shown in Fig. \ref{map}. The contours represent the geopotential and the vectors represent the wind. At the equator, the wind is more zonal, and the contours are more parallel to the equator. Away from the equator we can see cyclonic and anticyclonic vortices. In this figure, a wavenumber 2 pattern stands out, $i.e.$ the zonal wind alternates from eastward to westward twice in the domain. Consider one cycle of the pattern, that between $100^{o}$ and $280^{o}$ longitude. The maximum divergence of the zonal wind occurs at $180^{o}$. To the west of the maximum divergence, at about $160^{o}$, there are anticyclones to the north and the south, collocated with high geopotential anomalies. To the east of the divergence of the zonal wind, there are cyclones centered at $\sim 10^{o}$ latitude, together with low geopotential anomalies. This structure is referred to as the quadrupole vortex structure (\cite{MajdaStechmann2009}). The quadrupole vortices in our simulations are confined more closely to the equator than in the MJO as reported by \cite{Kiladis2005mjo}, where the off--equatorial vortices center at $\sim 20^o$ latitude.\\

\section{Proposed mechanism}

MJO-like signals have been simulated in our model. The simulated signal captures major features of the MJO, including propagation speed, horizontal scale, multi-scale structures and quadrupole vortex structures. Then we need to understand what the energy source is to maintain the long-lasting signal, and also need to address the eastward propagation mechanism. Figure \ref{finescale} shows high resolution Hovmoller diagrams. In this high resolution view, we are able to diagnose how convection is triggered, and how waves are excited. In Fig. \ref{finescale}a, there is a standing oscillation (SO) at a longitude of $\sim 206^{o}$. On day 163.75, there is a local minimum of $\phi$ at $206^{o}$. This triggers a convection event. A quarter cycle later, the convection reaches its maximum value, and a quarter cycle after that, phi reaches a local maximum. Such cycles repeat 5 times in Fig. \ref{finescale} with a period of 0.5 day, which is twice $\tau_c$. The amplitude of convective heating is small, so there is no nonlinearity due to advection. Linear waves are excited. However, due to the positive-only heating, nonlinearity has been introduced into this system. Net zonal divergence of u collocates with the SO event. The SO is a representation of selectively amplified waves.  Thus energy is introduced into this shallow water system. The SO can also be viewed as a standing wave, composed of interfering eastward and westward waves with similar speeds. In the power spectra, WIG and EIG waves are the only waves that propagate toward each other with same meridional structures and similar propagation speed. Therefore, we propose that the simulated MJO signal is an interfering pattern between WIG and EIG waves, which are excited by convection. The propagation speed of the MJO pattern is associated with the phase speed difference between WIG and EIG waves. \\

 Figure \ref{speed} shows how the simulation results matches the theoretically derived propagation speed. If the hypothesis is correct, the MJO propagation speed is one half of the phase speed difference between the WIG and EIG waves. In Eq. \eqref{trig}, the EIG phase speed is $c_1$, the WIG phase speed is $c_2$, and the cosine factor on the right is the standing oscillation. The sine factor on the right is the drift of the pattern at the MJO propagation speed.

  \begin{align}\label{trig} 
  \begin{split}
  sin[k(x-c_1t)] + sin[k(x+c_2t)] = 2sin[k(x-\frac{c_1-c_2}{2}t)]cos(k\frac{(c_1+c_2)t}{2})
  \end{split}
  \end{align}
 The thick black lines denote the theoretically derived MJO propagation speed from the dispersion relation of IG waves, and the markers represent the results from simulations with different convective timescale. The lower thick line is for the first meridional mode, and the higher thick line is for the third meridional mode. The abscissa is the corresponding frequency of IG waves. The theoretically derived speed does not fit the simulated results perfectly, especially for short convective timescales (converted into frequency). Increasing the convective timescale, the MJO propagation speed increases monotonically, and this behavior is qualitatively consistent with the theoretically predicted behavior. The marker starts to locate between the two thick lines when $\tau_c$ is longer than 0.6 day, $i.e.$, the simulated MJO signal propagates faster than the predicted speed from the first meridional mode, but slower than that of the third meridional mode. This suggests either that convection amplifies higher meridional modes at higher frequencies or that convection excites lower frequencies than $\frac{1}{2\tau_c}$. It is possible that due to the positive-only convective heating, waves with frequency lower than $\frac{1}{2\tau_c}$ will  be amplified.\\
 
As an interference pattern, the horizontal scale of the MJO will be associated with the bandwidth of the excited waves. Our simulations show that the bandwidth of waves is associated with how frequently the convection happens. If convection is very frequent, a wide band of waves will be excited. However, if there are only a few convection events during the simulation, only a narrow band of waves will be excited. Since the convection frequency per unit area $N$ is of order $\frac{3r}{q_o}$, decreasing $r$ will decrease $N$ if all the other parameters are fixed.  As a result, the MJO wavenumber decreases. Figure 7 shows the results of systematic experiments where the MJO scale increases as r increases. The amplitude of the convective forcing is small so that the waves are linear waves, and increasing the convective strength does the same job as decreasing $r$. \\

 \section{Discussion and Conclusions}
 
 In this paper, we have presented a shallow water model with triggered convection and simple radiation treatments. MJO-like signals are observed in our simulations. Our simulation results are robust over a wide parameter range, and do not depend on resolutions. We further propose that the MJO-like signal is an interference pattern of the WIG and EIG waves, whose frequency is set by the duration of individual convective events. We have tested this hypothesis by systematically varying the parameters $\tau_c$ and $N$. Both the propagation speed and the zonal wavenumber of the MJO-like signals respond to the parameter changes consistently with the theoretical predictions. Instead of the structure of the IG waves, the MJO-like signal exhibits the  quadrupole vortex structures. This might indicate that these off--equatorial vortices are responses to the large-scale envelope of convective heating near the equator, rather than free wave structures. Our simulation results suggest that the MJO perhaps is not a large-scale low-frequency wave in which, convection acts as QE adjustment. Small-scale high-frequency waves might be crucial. Therefore, in order to simulate the MJO, the behavior of moist convection in a short time period should be represented properly. \\
 
 We use a one-layer shallow water model without explicit moisture variables. However, it does not mean that we exclude the effect of moisture for the following reasons. First, triggered convection only occurs in the moist atmosphere, where conditional instability could exist. In a shallow water model, the geopotential $\phi$ is a measure of the static stability $N^2$ of the atmosphere. Low $\phi$ indicates the reduced $N^2$. Only if $N^2$ has been reduced to a critical value, in other words, CAPE has been accumulated by a certain amount, convection will be triggered. This mimics the process of conditional instability and triggered convection in a moist atmosphere with the minimum recipes. Second, we set the Kelvin wave speed close to the convectively coupled Kelvin wave speed, which is $\sim 16\ m\ s^{-1}$. \\
 
 Due to the simplicity of our model, we have to make assumptions about the MJO vertical structure and the Kelvin wave speed. Although these assumptions are consistent with observations, a complete theory will have to explain why the MJO has a first baroclinic structure, and why the Kelvin wave speed is $\sim$ 16 $m\ s^{-1}$. The next step is to use a 3D model to test our hypothesis. We do not need to assume the MJO vertical structure and the Kelvin wave speed in the 3D model. To keep the key features of our current model, we will implement triggered convection in the 3D model. In this model, CAPE will be accumulated until convective inhibition is close to zero. Then CAPE will be released.  \\

It is also worth to look for IG waves and their relations to convection and the MJO in observations to test our hypothesis. The correlation between the MJO and the IG waves has already been examined using observational datasets ($e.g.$ \cite{YangIngersoll2011}, \cite{YasunagaMapes2011}). Statistically significant correlations have been identified, but the IG waves can only explain very limited variance of the MJO. However, previous studies do not help to test our hypothesis. First, most of the previous studies used coarse temporal and spacial resolution datasets, so they cannot resolve high--frequency IG waves. Second, previous studies examined the correlations between the MJO and the WIG waves, and the MJO and the EIG waves separately. The interference pattern of the WIG and EIG waves are not even included. To test our hypothesis, one should first combine the WIG and EIG signals in a high--resolution dataset, and identify their interference patterns. Then examine the correlations between the MJO and this identified interference pattern. \\

\begin{acknowledgment} 
Da Yang was supported by the Earle C. Anthony Professor of Planetary Science Research Pool and the Division of Geological and Planetary Sciences Davidow Fund of the California Institute of Technology. He is currently supported by the Astronomy and Astrophysics Research Program of the National Science Foundation. We thank these organizations for their support. 
\end{acknowledgment}








\ifthenelse{\boolean{dc}}
{}
{\clearpage}
\bibliographystyle{ametsoc}
\bibliography{YangIngersollReference}

\clearpage

\begin{figure*}[htbp]
  \noindent\includegraphics[width=19pc,angle=0]{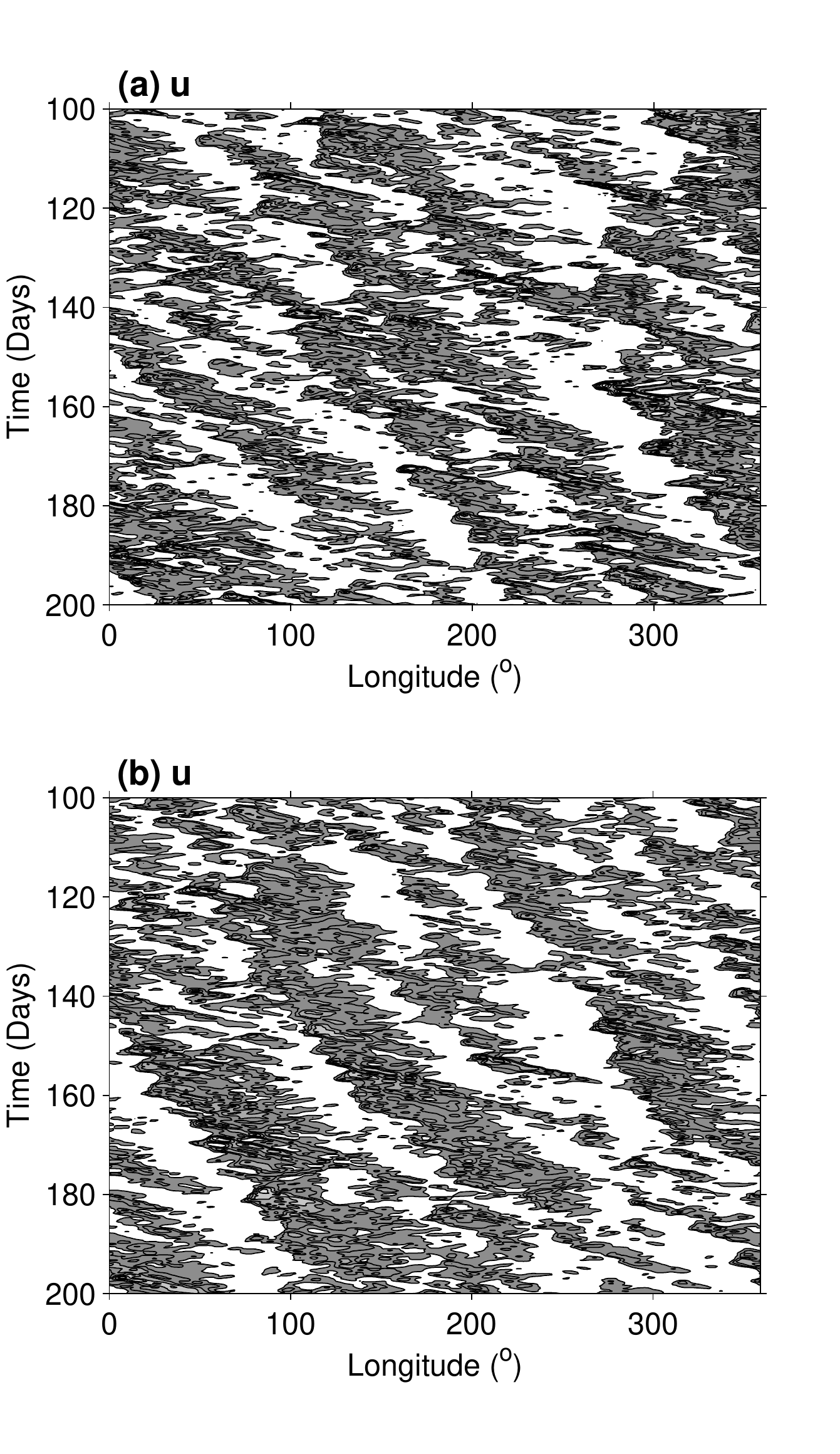}\\
  \caption{Hovmoller diagrams of zonal wind from (a) T85 simulation, (b) T170 simulation. The white represents low, and the black represents high.}\label{hdres}
\end{figure*}

\begin{figure*}[htbp]
  \noindent\includegraphics[width=19pc,angle=0]{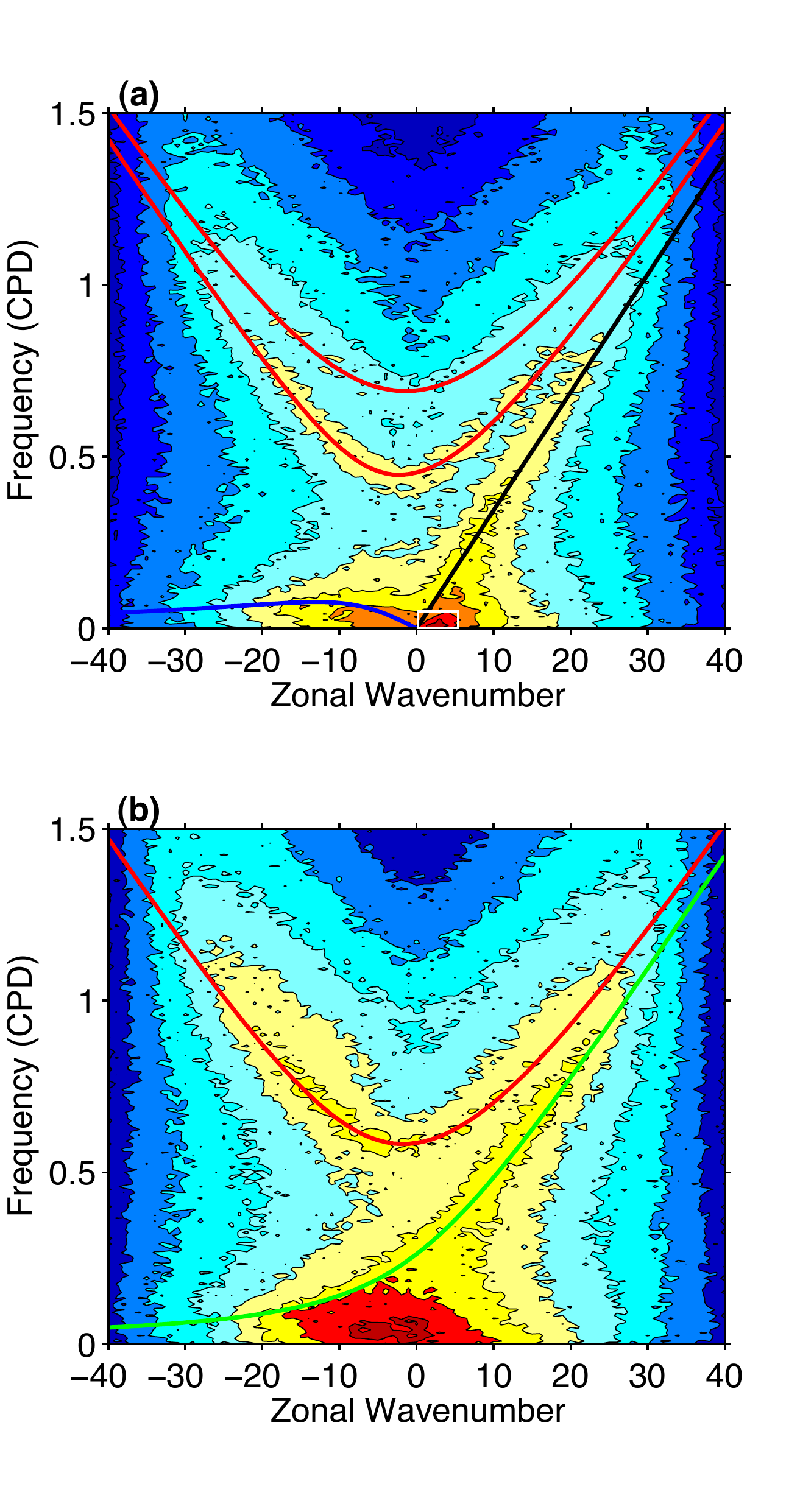}\\
  \caption{Zonal wavenumberÐfrequency power spectra of zonal wind from -$15^{o}$ to $15^{o}$ latitude. (a) is for the symmetric component, and (b) is for the anti-symmetric component. Red represents high power density, and blue represents low power density. Red, blue, black and green lines denote dispersion curves of IG, Rossby, Kelvin and MRG waves for different meridional modes.The white box ranges from 1 to 5 in wavenumber, and from $\frac{1}{60}$ to $\frac{1}{200}$ cycle per day (CPD) in frequency.}\label{spectra}
\end{figure*}

\begin{figure*}[htbp]
  \noindent\includegraphics[width=39pc,angle=0]{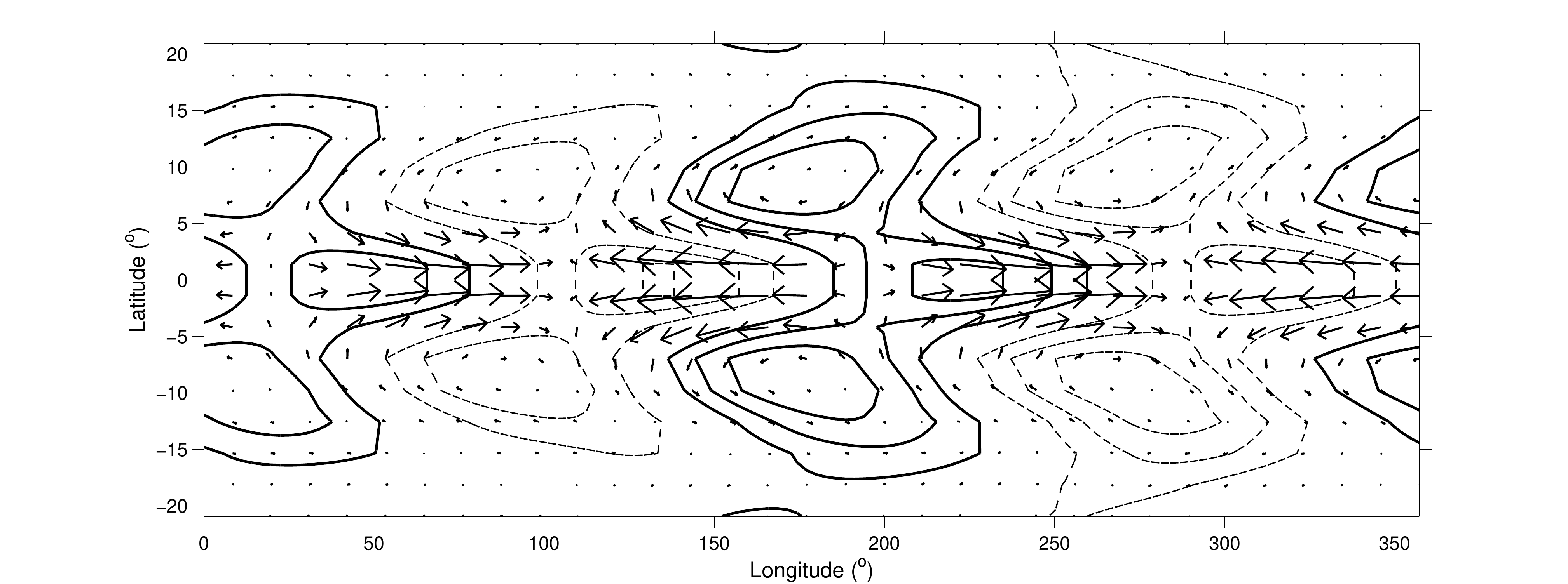}\\
  \caption{Longitude-latitude map of composites of the MJO-like signals. The arrows denote the wind field $\vec{V} = (u, v)$ , and the contours denote the geopotential. The thick solid (thin dashed) contours represent positive (negative) anomalies.}\label{map}
\end{figure*}

\begin{figure*}[htbp]
  \noindent\includegraphics[width=39pc,angle=0]{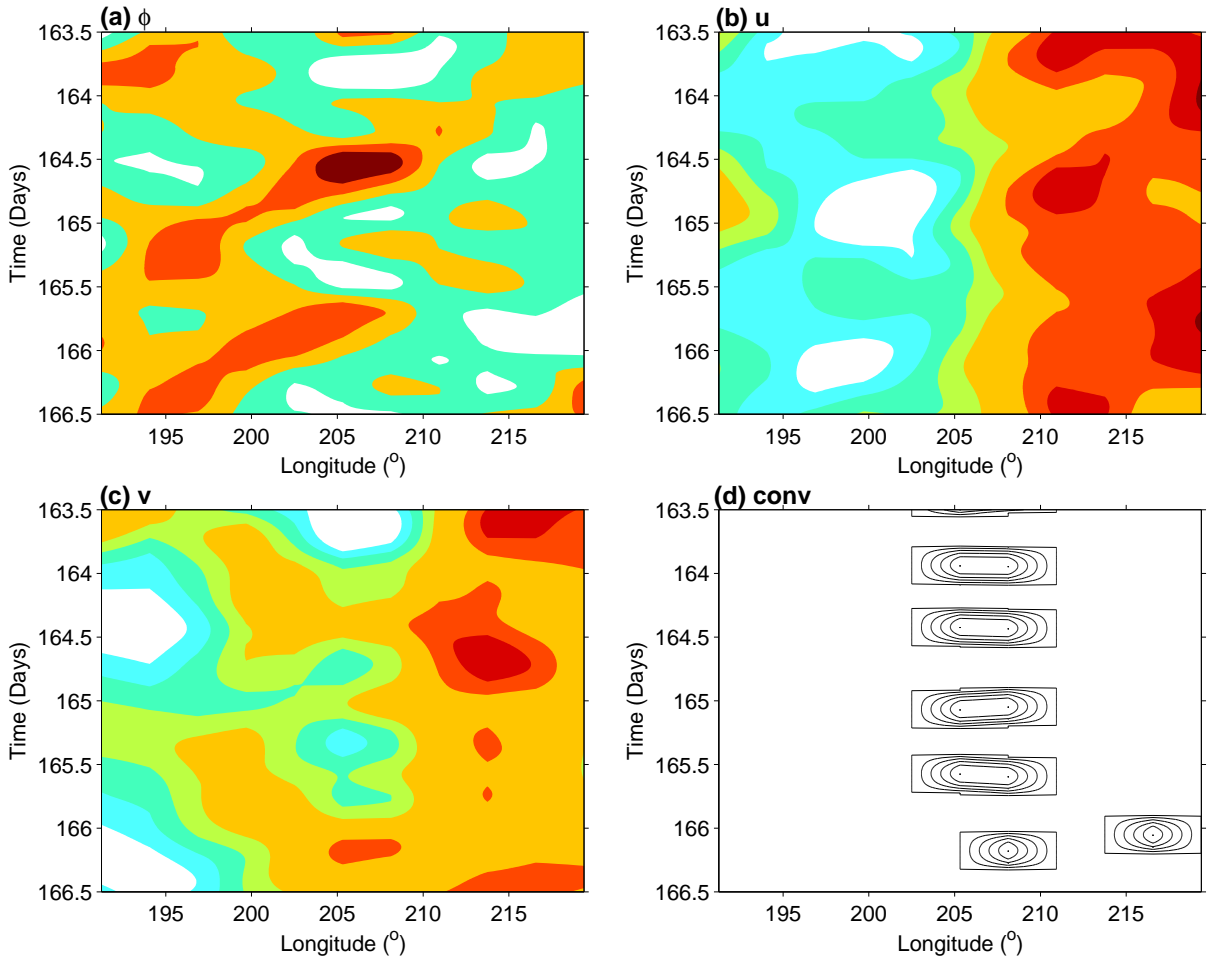}\\
  \caption{Hovmoller diagrams of the (a) geopotential, (b) zonal wind, (c) meridional wind and (d) convective heating of our shallow water simulation from -$15^{o}$ to $15^{o}$ latitude. The geopotential, zonal wind and convective heating are symmetric components about the equator, and the meridional wind is the anti-symmetric component. The blue contours denote low, and the red contours denote high.}\label{finescale}
\end{figure*}

\begin{figure*}[htbp]
  \noindent\includegraphics[width=19pc,angle=0]{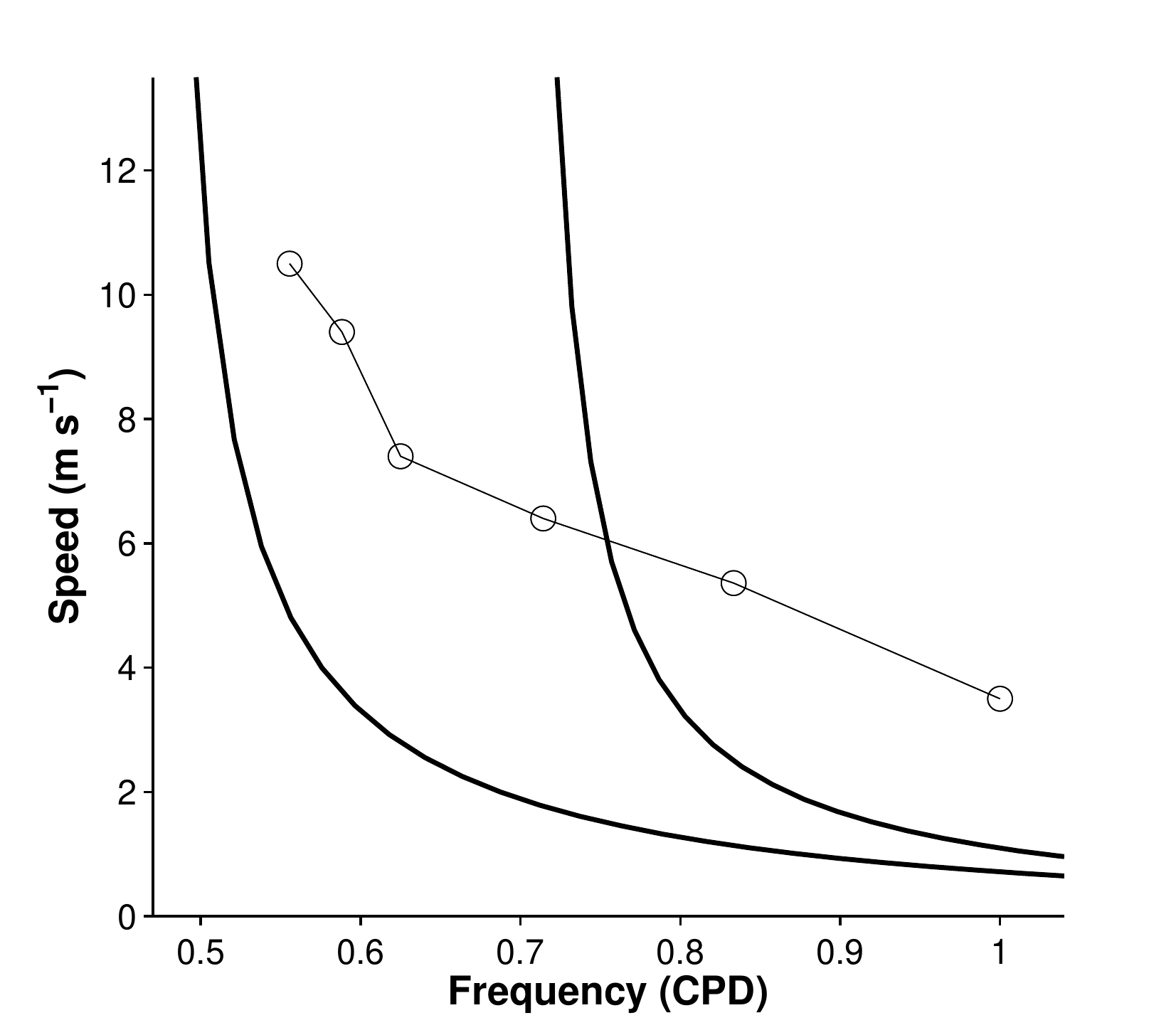}\\
  \caption{Propagation speed ($m\ s^{-1}$) versus frequency (CPD). The dark lines are derived from the dispersion relation of WIG and EIG waves. The lower one is for the first meridional mode, and the upper one is for the third meridional mode. The markers represent the simulation results for different convective timescales ($\tau_c$). We convert $\tau_c$ to frequency by using frequency  = $ \frac{1}{2\tau_c}$.}\label{speed}
\end{figure*}

\begin{figure*}[htbp]
  \noindent\includegraphics[width=19pc,angle=0]{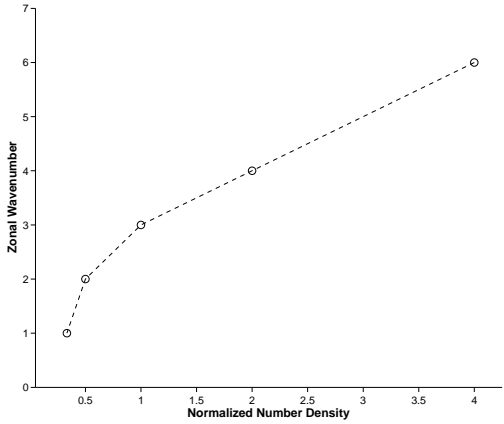}\\
  \caption{Zonal wavenumber versus normalized number density of convection $\widetilde{N}$, where $\widetilde{N} = \frac{N}{N_{o}}$. $N_o$ is given in Table \ref{t1}.}\label{k_sigma}
\end{figure*}

\begin{table}[t]
\caption{Parameter values in the control simulation.}\label{t1}
\begin{center}
\begin{tabular}{ccccrrcrc}
\hline\hline
$\phi_c$ $(m\ s^{-1})^2$ & $R \ (^{o})$ & $\tau_{c}$ (day) & $N$ ($m^{-2}\ s^{-1}$)  \\
\hline
250 & 3 & 0.25  & $1.12\times10^{-17}$   \\
\hline
\end{tabular}
\end{center}
\end{table}


%
\end{document}